\documentclass[12pt]{article}
\usepackage{amssymb,amsmath}
\begin{document}
\begin{flushright}
\vspace{ 1mm } hep-th/0207085
\\
FIAN/TD/06/12\\
June 2002\\
\end{flushright}
\vspace{10mm}
\begin{center}

{\Large \bf On vacuum-vacuum amplitude and Bogoliubov coefficients }\\

 \vspace{10mm} A.I.Nikishov ${}^\dag$\\

\vspace{3mm}{$\dag$\it I.E.Tamm Department of Theoretical Physics,
              Lebedev Physical Institute,\\
 Leninsky Prospect 53, 119991, Moscow, Russia\\
             e-mail: nikishov@lpi.ru}

    \vspace{2mm}

    \end{center}
\begin{abstract}
We consider the problem of fixing the phases of Bogoliubov coefficients
in Quantum Electrodynamics so that the vacuum -vacuum amplitude can be
expressed through them. For the constant electric field and particles with
spin 0 and 1/2 this is done starting from the definition of these coefficients.
Using the symmetry between electric and magnetic fields the result is
extended to the constant electromagnetic field. It turns out that for the case
 of a constant magnetic field it is necessary to distinguish the in- and
 out-states  although they differ only by a phase factor. For spin-1 particle
 with gyromagnetic ratio $g=2$ this approach fails and we reconsider
the problem by the proper-time method.
\end{abstract}

\section{Introduction}

\vskip-0.1cm
  Even if the electromagnetic field does not create pairs, virtual pairs lead
to the appearance of a phase in vacuum-vacuum amplitude. This makes it
necessary to distinguish the in- and out-solutions even when
it is commonly assumed that there is only one
complete set of solutions as, for example, in the case of a constant magnetic
field. Then in- and out-solutions differ only by a phase factor which is in
essence the Bogoliubov coefficient. The propagator  in
terms of in- and out-states  takes the same form as the one for pair creating
 fields.

We use the solutions with conserved quantum numbers and do not consider
radiation processes. Then the events in a cell with quantum numbers $n$
are independent of events in cells with different quantum numbers. In other
words we work in the diagonal representation.
The knowledge of Bogoliubov coefficients is sufficient for obtaining the
probability of any process in an external field (disregarding the radiation
 processes) [1-3]. But the real part of action integral $W$, defining the
 vacuum-vacuum
amplitude
$$
<0_{out}|0_{in}>=e^{iW},\quad W=\int d^4x{\cal L},       \eqno(1)
$$
is not directly expressed through Bogoliubov coefficients. At the same time some
effects connected with ${\rm ReW}$ are observable.
So the Lagrange function ${\cal L}$
of a slowly varying  field defines the dielectric permittivity  and
magnetic permeability of the field [4-5].

The Lagrange function of the constant electromagnetic field in one loop
 approximation was obtained in [6-8] and in two loop approximation in [9].
 Studying a model of particle production, B.DeWitt noted that ${\rm Re W}$
 can be
expressed through Bogoliubov coefficients with the natural choice of
 their phases [10]. Our purpose is to choose these phases in such
a way that ${\rm Re W}$ can be expressed through them. We show that for the
constant electric field and particles with spins 0 and 1/2 the natural choice
would be sufficient if it were not for the necessity to make renormalizations.
 For vector
boson with gyromagnetic ratio $g=2$ the situation is more complicated even
for the case of a constant electric field.

We note here that  transition
{\sl amplitude} for an electron to go from an in-state to out-state
turns out to be unity.  To show this we write down the Bogoliubov
transformations and the relation between $<0_{n\;out}|$ and  $<0_{n\;in}|$ [2]
($n$ is the set of quantum numbers)
$$
a_{n\;out}=c_{1n}a_{n\;in}-c_{2n}^*b_{n\;in}^+,
$$
$$
b_{n\;out}^+=c_{2n}a_{n\;in}+c_{1n}^*b_{n\;in}^+,
$$
 $$
<0_{n\;out}|=<0_{n\;in}|(c_{1n}^*-c_{2n}a_{n\;in}b_{n\;in}),\quad
 |c_{1n}|^2+|c_{2n}|^2=1.                                          \eqno(1a)
$$
Here $a_{n\;in}$  $(b_{n\;in}^+)$ is the particle (antiparticle) annihilation
(creation) operator, $a_{n\;in}|0_{n\;in}>=0$ and similarly for out-states.
$|0_{n\;in}>$ is the vacuum state in the cell with quantum number $n$. $c_{1n}$,
$c_{2n}$ are Bogoliubov coefficients and star means complex conjugation.

From the third relation in (1a) we have eq. (28) below and from the
first one
$$
a_{n\;in}^+=c_{1n}^{*-1}[a_{n\;out}^++c_{2n}b_{n\;in}].
$$
 Using this relation and anticommutator $\{a_{n'\;out},a_{n\;out}^+\}=
\delta_{n',n}$, we find [2]
$$
 <0_{n\;out}|a_{n\;out}a_{n\;in}^+|0_{n\;in}>=c_{1n}^{*-1}
<0_{n\;out}|0_{n\;in}>=1.                                        \eqno(1b)
$$
 The Pauli principle prohibits virtual pair
 creation in the state occupied by the electron. So even the phase of
scattering amplitude remains unchanged. In particular,
 for the constant
 magnetic field  $c_{2n}=0$ but we cannot assume $c_{1n}=1$
without violating eq. (1b) and eqs. (28-29) below because $W\ne0$ [4-5].
In other words, even when $c_{2n}=0$ the in- and out-vacuum are different.
(This is in contrast with the remark after eq. (15) in [10].)
So the Bogoliubov coefficient $c_{1n}$ have to be
coordinated with vacuum-vacuum amplitude. For the constant electromagnetic
field we represent the action integral $W=\int d^4x{\cal L}(x)=\sum_nW_n$ as
a sum over the set of quantum numbers  $n$. Then $W_n$ define (in general
complex) phase of Bogoliubov coefficient.

In Sec. 2 and 3, starting from the definition of Bogoliubov coefficients,
we consider the phase fixing for particles with spin 0 and 1/2 correspondingly.
In Sec. 4-6 we reconsider the problem by a more general proper-time
method for spins 0, 1/2 and 1.
\section{Scalar particle in the constant electromagnetic field}
For the case of set of wave functions with conserved quantum numbers $n$
 the Bogoliubov
 transformation have the form
$$
{}_+\psi_n=c_{1n}\;{}^+\psi_n+c_{2n}\;{}^-\psi_n,
$$
 $$
{}_-\psi_n=c_{2n}^*\;{}^+\psi_n+c_{1n}^*\;{}^-\psi_n;\\
 $$
$$
|c_{1n}|^2-|c_{2n}|^2=1.                                        \eqno(2)
$$
Here ${}_{+}\psi_n$ ($ {}^{+}\psi_n$) is the positive-frequency in- (out-)
solution and similarly for the negative-frequency states.

We are free to choose the phase of $c_{1n}$ by redefining $\psi_n$. Indeed, if
we substitute
$$
{}_{\pm}\psi_n=e^{\pm if}{}_{\pm}\psi_n^{new},\quad
{}^{\pm}\psi_n=e^{\mp if}{}^{\pm}\psi_n^{new},\quad
c_{1n}=e^{i2f}c_{1n}^{new},
$$
then eq. (2) and the propagator [2, 11]
$$
G_0(x,x')=i\sum_nc_{1n}^{*-1}\left\{\begin{array}{cc}
{}^+\psi_n(x){}_+\psi_n^*(x'),\quad t>t',\\
{}_-\psi_n(x){}^-\psi_n^*(x'),\quad t<t'
\end{array}\right.                                         \eqno(2a)
$$
 will have the same form in terms of redefined quantities.

 For definiteness we assume that the particle charge is $e'=-e, e=|e|$. 
For a constant electric field we have [2] ($n=(p_1,p_2,p_3),\; A_{\mu}=
-\delta_{\mu3}Et$)
$$
c_{1n}=\frac{\sqrt{2\pi}}{\Gamma(\frac12-i\varkappa)}\exp(-\frac{\pi\varkappa}
2+i\frac{\pi}4), \;c_{2n}=\exp(-\pi\varkappa-i\frac{\pi}2),  \quad
\varkappa=\frac{m^2+p_1^2+p_2^2}{2eE}.                       \eqno(3)
$$
We now note that in a weak electric field, $|c_{2n}|$ is exponentially small
and may be neglected. Then in- and out-states differ only by a phase factor.
The same should be true for the magnetic field where $c_{2n}=0$ exactly and
$\ln c_{1n}^*$ should be determined.

The probability amplitude that the vacuum in the state $n$ remains vacuum is [2]
$$
    <0_{n\;out}|0_{n\;in}>=c_{1n}^*{}^{-1}.                     \eqno(4)
$$
The total vacuum-vacuum amplitude is
$$
<0_{out}|0_{in}>=\prod\limits_nc_{1n}^*{}^{-1}=e^{iW_0},\;
W_0=\sum\limits_nW_{0n},\; W_{0n}=i\ln c_{1n}^*                 \eqno(5)
$$
Only ${\rm ReW_0}$ contain divergences and
as we shall see below, $c_{1n}^*$ in (4) and (5) should be replaced by
$ C_{1n}^{*ren}$. This is the renormalization of $c_{1n}^*$.
From (3) we have
$$
\ln c_{1n}^*=\frac12\ln2\pi-\frac{\pi\varkappa}2-\frac{i\pi}4-
\ln \Gamma(\frac12+i\varkappa).                                   \eqno(6)
$$
As shown in [2], the vacuum-vacuum probability $|<0_{out}|0_{in}>|^2$ obtained
 from (5) and (3) agrees with Schwinger result [8]. This means that
${\rm Im} W_0$ is given correctly by (5) and (3). To find ${\rm Re}W_0$, we
 first consider the
asymptotic representation, see eq. (1.3.12) in [12]
$$
\ln\Gamma(\frac12+i\varkappa)=i\varkappa[\ln(i\varkappa)-1]+\frac12\ln2\pi+
\sum\limits_{k=1}\frac{B_{2k}(\frac12)}{2k(2k-1)}(i\varkappa)^{1-2k}.  \eqno(7)
$$
(Letting  $k$ run to $\infty$, we may say that the r.h.s. of (7) in a
certain sense exactly represent the l.h.s.; the information encoded in the
r.h.s. can be decoded [13].) From (6) and (7) it follows
$$
\ln c_{1n}^*=-i\{\varkappa(\ln\varkappa-1)+\frac{\pi}4+\sum\limits_{k=1}
\frac{(-1)^kB_{2k}(\frac12)}{2k(2k-1)\varkappa^{2k-1}}\}.            \eqno(8)
$$
This asymptotic expansion contains only the imaginary part of $\ln c_{1n}^*$
or only real part of $W_{0n}$.
It is seen from (8) that as a first step we have to go from $\ln c_{1n}^*$
to
$$
\ln C_{1n}^*=\ln c_{1n}^*+i\{\varkappa(\ln\varkappa-1)+\frac{\pi}4\}  \eqno(9)
$$
in order to have $\ln C_{1n}^*\to0$ for $\varkappa\to\infty$ (i.e. for $E\to0$).
 Due to the necessity
of charge renormalization we have to make the second step and introduce
$$
\ln C_{1n}^{*ren}=\ln c_{1n}^*+i\{\varkappa(\ln \varkappa-1)+\frac{\pi}4+
\frac1{24\varkappa}\}.                                                 \eqno(10)
$$
In other words we include in $\ln C_{1n}^{*ren}$ also the term with $k=1$
in (8). Then we have the following asymptotic representation
$$
\ln C_{1n}^{*ren}=-i\sum\limits_{k=2}
\frac{(-1)^kB_{2k}(\frac12)}{2k(2k-1)\varkappa^{2k-1}}.    \eqno(11)
$$

Summing (11) over $n$ according to the rule
$$
\sum\limits_k\to\int\frac{d^3pL^3}{(2\pi)^3},\quad \int dp_3\to eET, \eqno(12)
$$
and making renormalization [8], we obtain the correct asymptotic
representation for
${\rm Re}{\cal L}_0$
$$
{\rm Re}{\cal L}_0=\frac12E^2+\frac{(eE)^2}{16\pi^2}\sum\limits_{k=2}
\frac{(-1)^kB_{2k}(\frac12)}{k(k-1)(2k-1)\varkappa_0^{2k-2}},\quad
\varkappa_0=\frac{m^2}{2eE}.                                       \eqno(13)
$$
To simplify formulas and minimize confusion with $T$ in eq.(50) we often
put $L=T=1$ in the expressions like (12). Besides, in the following we
 drop the Maxwell part of the Lagrangian ($\frac12E^2$ in this case).

 Now we show that the expression (9) can be brought to the form suggested
by the proper-time formalism:
$$
\ln C_{1n}^*\equiv\ln\sqrt{2\pi}+\eta(\ln\eta-1)-\ln\Gamma(\frac12+\eta)=
-F(\eta),
$$
$$
 F(\eta)=\frac12\int\limits_0^{\infty}\frac{d\theta}{\theta}
e^{-2\eta\theta}[\frac1{\sinh\theta}-\frac1{\theta}].\quad \eta=i\varkappa.
                                                                    \eqno(14)
$$
Differentiating (14) with respect to  $\eta$ and using eq. (2.4.225) in [16], we see that
the results on the left- and the right-hand sides coincide. Besides, both
 sides have the same asymptotic behavior for $\eta\to\infty$. So we have
$$
\ln C_{1n}^*=-\frac12\int_0^{\infty}\frac{ds}{s\sinh\theta}
e^{-is(m^2+p_{\perp}^2)}[1-\frac{\sinh\theta}{\theta}],\theta=eEs, p_{\perp}=
p_1^2+p_2^2.                                                    \eqno(15)
$$
Next, we note that the term $\frac i{24\kappa}$ in (10) can be written as
$$
\frac i{24\varkappa}=-\frac1{12}\int_0^{\infty}d\theta e^{-i2\varkappa\theta}.
                                                             \eqno(16)
$$
Hence
$$
\ln C_{1n}^{*ren}=-\frac12\int_0^{\infty}\frac{ds}{s\sinh\theta}
e^{-is(m^2+p_{\perp}^2)}R(\theta),\quad
R(\theta)=1-\sinh\theta\left(\frac1{\theta}-\frac{\theta}6\right). \eqno(17)
$$
Here $R(\theta)$ is a "regularizer." It is independent of quantum numbers $n$
and is the same as in the proper-time representation of the Lagrange function
[8].

Now we consider the case when there is a constant magnetic field collinear
with the constant electric field. Then
$$
\ln C_{1n}^{*ren}(E,H)=-\frac12\int_0^{\infty}\frac{ds}{s\sinh\theta}
e^{-is(m^2+eH(2l+1))}R(\theta, \tau), \tau=eHs, l=0,1,\cdots   \eqno(18)
$$
and we assume that $R(\theta,\tau)$  may be obtained by the same reasoning
as in [8] (or simply taken from [8])
$$
R(\theta,\tau)=1-\sinh\theta \sin\tau\left(\frac1{\theta\tau}+
\frac16\frac{H^2-E^2}
{EH}\right),\quad\tau=eHs,\quad \theta=eEs.                \eqno(19)
$$

Integrating over $p_3$, we get (see (12) with $T=1$)
$$
\int dp_3\ln C_{1n}^{*ren}(E,H)=-\frac12eE
\int_0^{\infty}\frac{ds}{s\sinh\theta}
e^{-is(m^2+eH(2l+1))}R(\theta, \tau).   \eqno(20)
$$
 In this expression we can turn off the electric field
$$
\int dp_3\ln C_{1n}^{*ren}(E=0,H)=-\frac12
\int_0^{\infty}\frac{ds}{s^2}
e^{-is(m^2+eH(2l+1))}R(0,\tau),
$$
  $$
R(0,\tau)=1-\sin\tau\left(\frac1{\tau}+\frac{\tau}6\right).          \eqno(21)
$$
To remove the integration over $p_3$, we write the factor $s^{-2}$ as
$s^{-3/2}s^{-1/2}$ and note that  $1/\sqrt s$
 should be due the integration over $p_3$:
$$
\frac1{\sqrt s}=\frac{e^{i\pi/4}}{\sqrt\pi}
\int_0^{\infty}dp_3e^{-isp_3^2}.                               \eqno(22)
$$
So
$$
\ln C_{1n}^{*ren}(E=0,H)=-\frac{e^{i\pi/4}}{2\sqrt\pi}
\int_0^{\infty}\frac{ds}{s^{3/2}}
e^{-is(m^2+eH(2l+1)+p_3^2)}R(0,\tau).                                 \eqno(23)
$$
(Substituting $s\to-it$ we see that expression (23) is purely imaginary.)
From here, or from (21) we obtain in agreement with [8, 9]
$$
i\sum_n\ln C_{1n}^{*ren}(E=0,H)=i\int\frac{dp_2}{2\pi}\int\frac{dp_3}{2\pi}
\sum_{l=0}^{\infty}\ln C_{1n}^{*ren}(E=0,H)={\cal L}_0=
$$
$$
-\frac{eH}{16\pi^2}
\int_0^{\infty}\frac{ds}{s^2\sin\tau}e^{-ism^2}R(0,\tau),\quad( L=T=1)
                                                               \eqno(24)
$$
Relation (39) was used here and the sum
over $l$ was performed with the help of formula
$$
\sum_{l=0}^{\infty}e^{-iseH(2l+1)}=\frac1{2i\sin(eHs)}.    \eqno(25)
$$

\section{Electron in the constant electromagnetic field}
The Bogoliubov transformation has the form
$$
{}_+\psi_n=c_{1n}\;{}^+\psi_n+c_{2n}\;{}^-\psi_n,
$$
 $$
{}_-\psi_n=-c_{2n}^*\;{}^+\psi_n+c_{1n}^*\;{}^-\psi_n;\\
 $$
$$
|c_{1n}|^2+|c_{2n}|^2=1.                                         \eqno(26)
$$
For the constant electric field we have
$$
c_{1n}^*=-i\sqrt{\frac{2\pi}{\varkappa}}\frac{e^{-\frac{\pi\varkappa}{2}}}
{\Gamma(i\varkappa)},\quad c_{2n}=e^{-\pi\varkappa},\quad
 n=(p_1,p_2,p_3,r).                                                 \eqno(27)
$$
These Bogoliubov coefficients are independent of the spin state index
$r=1,2$.

As in the scalar case, we start with [2]
$$
    <0_{n\;out}|0_{n\;in}>=c_{1n}^*,                             \eqno(28)
$$
and
$$
<0_{out}|0_{in}>=\prod\limits_nc_{1n}^*=e^{iW_{1/2}},\;
W_{1/2}=\sum\limits_nW_{1/2;n},\; W_{1/2;n}=-i\ln c_{1n}^*         \eqno(29)
$$
From (27) we have
$$
\ln c_{1n}^*=-\frac{i\pi}{2}+\frac12\ln\frac{2\pi}{\varkappa}-
\frac{\pi\varkappa}{2}
-\ln\Gamma(i\varkappa).                  \eqno(30)
$$
The asymptotic expansion for $\Gamma(i\varkappa)$ is, see eq.(8.344) in [14]
or eq. (6.1.40.) in [15]
$$
\ln\Gamma(i\varkappa)=(i\varkappa-\frac12)\ln(i\varkappa)-i\varkappa
+\frac12\ln2\pi+
i\sum\limits_{k=1}(-1)^k\frac{B_{2k}}{2k(2k-1)}(\varkappa)^{1-2k}.  \eqno(31)
$$
From (30) and (31) we obtain
$$
\ln C_{1n}^*\equiv\ln c_{1n}^*+i(\varkappa\ln\varkappa-\varkappa+\frac{\pi}4)=
-i\sum\limits_{k=1}(-1)^k\frac{B_{2k}}{2k(2k-1)}(\varkappa)^{1-2k},  \eqno(32)
$$
$$
\ln C_{1n}^{*ren}\equiv \ln C_{1n}^*-\frac{i}{12\varkappa}=
-i\sum\limits_{k=2}(-1)^k\frac{B_{2k}}{2k(2k-1)}(\varkappa)^{1-2k}.  \eqno(33)
$$

 As in the scalar case we find
$$
\ln C_{1n}^*=-\frac12\int_0^{\infty}\frac{dx}{x}e^{-i2\varkappa x}
(\coth x-\frac1x),                                               \eqno(34)
$$
$$
\ln C_{1n}^{*ren}=-\frac12\int_0^{\infty}\frac{dx}{x}e^{-i2\varkappa x}
\coth x[1-\tanh x(\frac1x+\frac x3)],                            \eqno(35)
$$
Eq. (2.4.22.6) in [16] was used to verify (34), cf. the text before eq.(15).

The generalization of (35) for the presence of a constant magnetic field
is straightforward. We rewrite it in the form ($x=\theta=eEs$)
$$
\ln C_{1n}^{*ren}(E,H)=-\frac12\int_0^{\infty}\frac{d\theta}{\theta}
e^{-is(m^2+eH2l)}\coth\theta R(\theta,\tau),
$$
 $$
n=(p_1,p_2,p_3,r);\; l=l_{min},l_{min}+1,\cdots,\; \; l_{min}=0\quad
{\rm for} \; r=1,\; l_{min}=1 \quad
{\rm for}\; r=2.                                                    \eqno(36)
 $$
$R(\theta,\tau)$ may be taken from the Lagrange function [8, 9] $(\tau=eHs)$
$$
 R(\theta,\tau)=1-\tan\tau\tanh\theta(\frac1{\theta\tau}+\frac{E^2-H^2}{3EH}).
                                                                   \eqno(37)
$$
Integrating over $p_3$ with the help the second equation in (12) we find
$$
\int\frac{dp_3}{2\pi}\ln C_{1n}^{*ren}=-\frac{eE}{4\pi}
\int_0^{\infty}\frac{d\theta}{\theta}
e^{-is(m^2+eH2l)}\coth\theta R(\theta,\tau).                    \eqno(38)
$$
The subsequent integration over $p_2$ is performed according to formula similar
to (12) [2]
$$
\int dp_2=eHL.                                                         \eqno(39)
$$
To sum up over $r$ and $l$ in (36), we use the formula obtainable from (25)
$$
\sum_{r=1}^2\sum_{l_{min}}^{\infty}e^{-is2eHl}=-i\cot(eHs).    \eqno(40)
$$
So in agreement  with the Lagrange function for the constant electromagnetic
field [8, 9] we have
$$
\sum_n\ln C_{1n}^{*ren}=i\frac{e^2EH}{8\pi^2}\int_0^{\infty}
\frac{d\theta}{\theta}
e^{-ism^2}\coth\theta\cot\tau R(\theta,\tau),\quad (L=T=1).       \eqno(41)
$$

Now returning to (38), we can switch off the electric field
$$
\int\frac{dp_3}{2\pi}\ln C_{1n}^{*ren}=-\frac1{4\pi}
\int_0^{\infty}\frac{ds}{s^2}
e^{-is(m^2+eH2l)}R(0,\tau),
$$
 $$
R(0,\tau)=1-\tan\tau(\frac1{\tau}-\frac{\tau}3);                 \eqno(42)
$$
$l$ are given in (36).
Using (22) we obtain as in the scalar case
$$
\ln C_{1n}^{*ren}(E=0,H)=-\frac{e^{i\pi/4}}{2\sqrt{\pi}}
\int_0^{\infty}\frac{ds}{s^{3/2}}
e^{-is(m^2+p_3^2+eH2l)} R(0,\tau),
$$
 $$
n=(p_1,p_2,p_3,r);\quad l=0,1,2,\cdots\;{\rm for}\; r=1;\quad l=1,2,\cdots \;
{\rm for}\; r=2.                                                    \eqno(43)
 $$

In the subsequent Sections we give the heuristic derivation
 of $\ln C_{1n}^{*ren}$ not
resorting to $c_{1n}^*$, but using the proper-time method The main problem
 here is due to the necessity
to make renormalizations. We know how to renormalize ${\cal L}$ as a
whole, but we have to renormalize a contribution to it
from a particular state $n$. To do this we assume as before that the
regularizer does not depend on $n$.
\section{Scalar particle}
We take the vector-potential of a constant electromagnetic field in the form
$$
A_{\mu}=\delta_{\mu2}Hx_1-\delta_{\mu3}Et,  \eqno(44)
$$
but start with the  particle in a constant magnetic field, $E=0$
in (44). The
propagator with coinciding $x$ and $x'$ has the form (see for example [11])
$$
G_0(x,x|E=0,H)=i\sqrt\frac{eH}{\pi}\sum_{l=0}^{\infty}
\int_{-\infty}^{\infty}
\frac{dp_2}{2\pi}\int_{-\infty}^{\infty}\frac{dp_3}{2\pi}
\int_{-\infty}^{\infty}\frac{dp^0}{2\pi}\times
$$
$$
\int_0^{\infty}ds
\frac{D_l^2(\zeta)}{l!}\exp\{-is[m^2+eH(2l+1)+p_3^2-p_0^2]\},\quad
\zeta=\sqrt{2eH}(x_1+\frac{p_2}{eH}).                            \eqno(45)
$$
According to (1) we have to integrate ${\cal L}_0$ and hence $G_0(x,x)$ over
$d^4x$.  The integration over $x_1$ is done with the help of formula
$$
\int\limits_{-\infty}^{\infty}d\zeta D_l^2(\zeta)=\sqrt{2\pi}l!,\quad {\rm or}
\int\limits_{-\infty}^{\infty}dx_1D_l^2(\zeta)=
\left(\frac{\pi}{eH}\right)^{1/2}l!.
                                                         \eqno(46)
$$
Integrating over $p^0$ and $x_1$, we obtain
$$
\int_{-\infty}^{\infty}dx_1G_0(x,x)=\frac{\exp[i\frac{3\pi}4]}{2\sqrt{\pi}}
 \int_{-\infty}^{\infty}\frac{dp_2}{2\pi}
\int_{-\infty}^{\infty}\frac{dp_3}{2\pi}
\sum_{l=0}^{\infty}
\int_0^{\infty}\frac{ds}{\sqrt s}e^{-is[m^2+eH(2l+1)+p_3^2]}.    \eqno(47)
$$
As noted in [3] (see eq. (2.12) there), it follows from Schwinger results [8]
that for scalar particle (boson)
$$
-i\frac{\partial{W}_b}{\partial m^2}=\int d^4xG_b(x,x),\quad{\rm or}\quad
 {W}_b=
-i\int_{m^2}^{\infty}d\tilde m^2\int d^4xG_b(x,x|\tilde m^2).
                                                                  \eqno(48)
$$
This means that ${\cal L}_0$ can be obtained from (47) by
 inserting $-1/s$ in the
integrand. Inserting also the regularizer from (21), we get
$$
iW_0(E=0,H)=i{\cal L}_0(E=0,H)=\frac{\exp[i\frac{\pi}4]}{2\sqrt{\pi}}
 \int_{-\infty}^{\infty}\frac{dp_2}{2\pi}\int_{-\infty}^{\infty}
\frac{dp_3}{2\pi}\sum_{l=0}^{\infty}
\int_0^{\infty}\frac{ds}{ s^{3/2}}\times
$$
$$
e^{-is[m^2+eH(2l+1)+p_3^2]}R(0,\tau)=
- \int_{-\infty}^{\infty}\frac{dp_2}{2\pi}\int_{-\infty}^{\infty}
\frac{dp_3}{2\pi}\sum_{l=0}^{\infty}
\ln C_{1n}^{*ren}.\quad
(L=T=1)                                                              \eqno(49)
$$

Now for the constant electromagnetic field, described by the vector-potential
(44), we substitute in (2a) the expressions for the wave functions
(see [2] with the modifications for $e'=-e=-|e|$) and use relation (93)
in [11] (or relation similar to (96) below). Than we find
$$
G_0(x,x|E,H)=\frac{e^{i3\pi/4}
}{2\sqrt{\pi eE}}\int\limits_{-\infty}^{\infty} \frac{dp_2}{2\pi
}\int\limits_{-\infty}^{\infty}\frac{dp_3}{2\pi}\sum_{l=0}^{\infty}
\left(\frac{eH}{\pi}\right)^{1/2}\frac{D_l^2}{l!}
\sqrt2\int\limits_0^{\infty}\frac{d\theta}{\sqrt{\sinh2\theta}}\times
$$
$$
\exp[-i2\varkappa\theta-i\frac{T^2}{2\coth\theta}],\quad
\theta=eEs,\quad T=\sqrt{2eE}(t-\frac{p_3}{eE}).                 \eqno(50)
$$
 Integrating over $x_1$ (see (46)) and $t$, we get
$$
\int\limits_{-\infty}^{\infty}dx_1
\int\limits_{-\infty}^{\infty}dtG_0(x,x|E,H)=
$$
$$
\frac i2\int_{-\infty}^{\infty}\frac{dp_2}{2\pi}
\int_{-\infty}^{\infty}\frac{dp_3}{2\pi}\sum_{l=0}^{\infty}
\int\limits_0^{\infty}\frac{ds}{\sinh(eEs)}\exp[-is(m^2+eH(2l+1))].   \eqno(51)
$$
Passing from $G_0(x,x)$ to ${\cal L}_0$ is realized by inserting the factor
$(-1/s)$ in the integrand in (51). Inserting also the regularizer
 $R(\tau,\theta)$, see eq. (19), we obtain
$$
W_0(E,H)=i\int\limits_{-\infty}^{\infty}\frac{dp_2}{2\pi}
 \int\limits_{-\infty}^{\infty}\frac{dp_3}{2\pi}
\sum_{l=0}^{\infty}\ln C_{1n}^{*ren}
$$
$$
 =-\frac i2\int\limits_{-\infty}^{\infty}\frac{dp_2}{2\pi}
 \int\limits_{-\infty}^{\infty}\frac{dp_3}{2\pi}
\sum_{l=0}^{\infty}
\int\limits_0^{\infty}\frac{ds}{s\sinh\theta}
\exp\{-is[m^2+eH(2l+1)]\}R(\tau,\theta).                   \eqno(52)
$$

\section{Spinor particle}
First we consider the electron in the constant magnetic field, $E=0$ in
(44). The squared Dirac equation can be brought to the form
$$
\{\frac{d^2}{d\zeta^2}-\frac{\zeta^2}4+\frac{p_0^2-p_3^2}{2eH}
-\frac12\Sigma_3\}Z=0,\quad
\Sigma_3=\left(\begin{array}{cc}
\sigma_3& 0\\
0& \sigma_3
\end{array}\right).                                         \eqno(53)
$$
Here $\zeta$ is the same as in (45). We see that $Z$ can be written as follows
$$
Z={\rm diag}(f_1,f_2,f_1,f_2)e^{i(p_2x_2+p_3x_3-p^0t)}   \eqno(54)
$$
and $f_1$, $f_2$ have to satisfy the equation
$$
\{\frac{d^2}{d\zeta^2}-\frac{\zeta^2}4+\frac{p_0^2-p_3^2}{2eH}
\mp\frac12\}f_{1,2}=0.                                                \eqno(55)
$$
We choose $f_1=D_{l-1}(\zeta)$, $f_2=D_l(\zeta)$ in order that $p_{\perp}^2=
2eHl$ in both cases.
The solutions of the Dirac equation are obtained as the columns of the
matrix  [2]
$$
(m-i\hat\Pi)Z,\quad\hat\Pi=\gamma_{\mu}\Pi_{\mu},\quad
 \Pi_{\mu}=-i\partial_{\mu}+eA_{\mu}.                          \eqno(56)
$$

 Using the $\gamma-$matrices in the standard representation [4] we have
$$
(m-i\hat\Pi)=\left(\begin{array}{cccc}
m+\Pi^0&0&-\Pi_3&-\Pi_1+i\Pi_2\\
0&m+\Pi^0&-\Pi_1-i\Pi_2&\Pi_3\\
\Pi_3&\Pi_1-i\Pi_2&m-\Pi^0&0\\
\Pi_1+i\Pi_2&-\Pi_3&0&m-\Pi^0
\end{array}\right).                                              \eqno(57)
$$
In terms of $\zeta$ we get
$$
\Pi_1+i\Pi_2=-i\sqrt{2eH}\left(\frac{d}{d\zeta}-\frac{\zeta}2\right),\quad
\Pi_1-i\Pi_2=-i\sqrt{2eH}\left(\frac{d}{d\zeta}+\frac{\zeta}2\right). \eqno(58)
$$
Using also the relations
$$
\left(\frac{d}{d\zeta}+\frac{\zeta}2\right)D_l(\zeta)=lD_{l-1}(\zeta),
\left(\frac{d}{d\zeta}-\frac{\zeta}2\right)D_l(\zeta)=-D_{l+1}(\zeta),\eqno(59)
$$
we find (the exponential factor in (54) is dropped here for brevity)
$$
(m-i\hat\Pi)Z=
$$
$$
\left(\begin{array}{cccc}
(m+p^0)D_{l-1}(\zeta)&0&-p_3D_{l-1}(\zeta)&il\sqrt{2eH}D_{l-1}(\zeta)\\
0&(m+p^0)D_l(\zeta)&-i\sqrt{2eH}D_l(\zeta)&p_3D_l(\zeta)\\
p_3D_{l-1}(\zeta)&-il\sqrt{2eH}D_{l-1}(\zeta)&(m-p^0)D_{l-1}(\zeta)&0\\
i\sqrt{2eH}D_l(\zeta)&-p_3D_l(\zeta)&0&(m-p^0)D_l(\zeta)
\end{array}\right).                                              \eqno(60)
$$
Choosing the second and the first columns as $\psi_1$ and $\psi_2$ (subscripts
1 and 2 indicate spin states) and normalizing them, we get

$$
{}_+\psi_1=N_n\begin{bmatrix}
0\\
(m+p^0)D_l(\zeta)\\
-il\sqrt{2eH}D_{l-1}(\zeta)\\
-p_3D_l(\zeta)
\end{bmatrix}e^{iq\cdot x},\quad N_n=\left(\frac{eH}{\pi}\right)^{1/4}
\sqrt{\frac1{2p^0(p^0+m)l!}},
$$
 $$
p^0=\sqrt{m^2+2eHl+p_3^2},\quad q\cdot x=p_2x_2+p_3x_3-p^0t,\;
$$
$$
n=(p_2,p_3,l,r),\quad \zeta=\sqrt{2eH}(x_1+\frac{p_2}{eH}),  \eqno(61)
$$
$$
{}_+\psi_2=N_n\sqrt l\begin{bmatrix}
(m+p^0)D_{l-1}(\zeta)\\
0\\
p_3D_{l-1}(\zeta)\\
i\sqrt{2eH}D_{l-1}(\zeta)
\end{bmatrix}e^{iq\cdot x},\quad l=0,1,2,\cdots   \eqno(62)
$$
As seen from (62) in this state $l$ begins actually from unity. The
 negative-frequency solutions ${}_-\psi_n$ are obtained from (61-62) by
substitution $q\to-q$. We note here that eqs. (61-62) differ from eq.
(10.5.9) in [4] because there the authors assumed the charge of a spinor
particle to be positive.

Having obtained the wave functions, we are going to find the contribution
to ${\cal L}_{1/2}$ from each state $\psi_n$. For the field which does
 not create pairs, the propagator has the usual form
$$
G_{1/2}(x,x')=i\sum_n\left\{\begin{array}{cc}
{}_+\psi_n(x){}_+\bar\psi_n(x'),\quad t>t',\\
-{}_-\psi_n(x){}_-\bar\psi_n(x'),\quad t<t,
\end{array}\right. \quad \bar\psi_n=\psi^*_n\beta.       \eqno(63)
$$
In the standard representation
$$
\beta=\left(\begin{array}{cc}
{\rm I}&0\\
0&-{\rm I}
\end{array}\right),\quad {\rm I}=\left(\begin{array}{cc}
1&0\\
0&1
\end{array}\right).                                  \eqno(64)
$$
From (61) and (64) we find
$$
{\rm tr}{}_+\psi_1(x){}_+\bar\psi_1(x)=N_n^2\{[(m+p^0)^2-p_3^2]D_l^2(\zeta)-
2eHl^2D_{l-1}^2(\zeta)\}.                               \eqno(65)
$$
Integrating over $x_1$, we obtain, see (46)
$$
\int\limits_{-\infty}^{\infty}dx{\rm tr}{}_+\psi_1(x){}_+\bar\psi_1(x)=
\frac m{p^0},\quad p^0=\sqrt{m^2+2eHl+p_3^2},\quad l=0,1,\cdots     \eqno(66)
$$
Similarly, from (62) we get
$$
\int\limits_{-\infty}^{\infty}dx{\rm tr}{}_+\psi_2(x){}_+\bar\psi_2(x)=
\frac m{p^0},\quad  l=1,2,\cdots                                  \eqno(67)
$$

For the negative-frequency states we have to substitute $p^0\to-p^0$. Now
we can write
$$
\frac1{|p^0|}=\frac{e^{\frac{i\pi}4}}{\sqrt{\pi}}\int\limits_0^{\infty}
\frac{ds}{\sqrt s}e^{-is(m^2+2eHl+p_3^2)},                           \eqno(68)
$$
accommodating both lines in (63).
Thus, from (63) and (66-68) we have
$$
\int\limits_{-\infty}^{\infty}dx_1G_{1/2}(x,x)=\sum_n
\frac{e^{\frac{3i\pi}4}}{\sqrt{\pi}}m\int\limits_0^{\infty}
\frac{ds}{\sqrt s}e^{-is(m^2+2eHl+p_3^2)},                          \eqno(69)
$$
where $l=0,1,\cdots$ for $r=1$ and $l=1,2,\cdots$ for $r=2$.
Next we use the analog of (48) for the electron
$$
 W_{1/2}=
i\int_{m}^{\infty}d\tilde m{\rm Tr}G_{1/2}(x,x|\tilde m).             \eqno(70)
$$
Here {\rm Tr} means the integration over $d^4x$ and the trace over spin
 indexes;
as above we put $VT=1$. Since
$$
i\int\limits_m^{\infty}d\tilde m\tilde me^{-is\tilde m^2}=
\frac{e^{-ism^2}}{2s},
                                                                  \eqno(70a)
$$
we see that $W_{1/2}$ may be obtained from (69) by inserting the factor
$(\frac1{2ms})$ into the integrand. So we find
$$
{\cal L}_{1/2}=\sum_n
\frac{e^{\frac{3i\pi}4}}{2\sqrt{\pi}}\int\limits_0^{\infty}
\frac{ds}{s^{3/2}}e^{-i(m^2+2eHl+p_3^2)}R(0,\tau).             \eqno(71)
$$
This is in agreement with (43) and (29). To check this result,
 we integrate over
$\frac{dp_2}{2\pi}$ with the help of (39), over $\frac{dp_3}{2\pi}$ with
 the help of (22) and use (40). Then, as expected, we get
$$
{\cal L}_{1/2}(E=0,H)=\frac{eH}{8\pi^2}\int\limits_0^{\infty}\frac{ds}{s^2}
e^{-ism^2}\cot\tau R(0,\tau),                                      \eqno(72)
$$
see eq. (47) in Ch. 1 in the last Ref. in [9] for $E=0$.

Going over to the constant electromagnetic field described by vector-potential
(44), we use $\gamma-$matrices in the spinor representation because  then
both $\alpha_3$ and $\Sigma_3$ are diagonal. The squared Dirac equation
has the form
$$
(\Pi^2+m^2+g)Z=0,\quad g=e\left(\begin{array}{cc}
(H-iE)\sigma_3&0\\
0&(H+iE)\sigma_3
\end{array}\right),                                             \eqno(73)
$$
$\Pi_{\mu}$ is defined in (56).
So
$$
Z={\rm diag}(f_1,f_2,f_3,f_4)e^{i(p_2x_2+p_3x_3)}.                                  \eqno(74)
$$

In terms of $\zeta$ and $T$ (see (45) and (50)) we obtain the equation
for $f_1$ and $f_2$
$$
\{2eH[-\frac{\partial^2}{\partial\zeta^2}+\frac{\zeta^2}4\pm\frac12]+
2eE[\frac{\partial^2}{\partial T^2}+\frac{T^2}4\mp\frac i2]+m^2\}f_{1,2}=0.
                                                                   \eqno(75)
$$
Similarly, for $f_3$ and $f_4$
$$
\{2eH[-\frac{\partial^2}{\partial\zeta^2}+\frac{\zeta^2}4\pm\frac12]+
2eE[\frac{\partial^2}{\partial T^2}+\frac{T^2}4\pm\frac i2]+m^2\}f_{3,4}=0.
                                                                   \eqno(76)
$$
From these equations we obtain
$$
{}^+Z={\rm diag}\{D_{l-1}(\zeta)D_{-i\varkappa-1}(\chi),
D_l(\zeta)D_{-i\varkappa}(\chi),D_{l-1}(\zeta)D_{-i\varkappa}(\chi),
D_l(\zeta)D_{-i\varkappa-1}(\chi)\}\times
$$
$$
e^{i(p_2x_2+p_3x_3)},\quad \chi=e^{\frac{i\pi}4}T. \eqno(77)
$$
Solutions of the Dirac equation with $\gamma-$matrices in spinor
 representation are obtained as the columns of the matrix
$$
(m-i\hat\Pi)Z=\left(\begin{array}{cccc}
m&0&\Pi^0+\Pi_3&\Pi_1-i\Pi_2\\
0&m&\Pi_1+i\Pi_2&\Pi^0-\Pi_3\\
\Pi^0-\Pi_3&-\Pi_1+i\Pi_2&m&0\\
-\Pi_1-i\Pi_2&\Pi^0+\Pi_3&0&m
\end{array}\right)Z.                          \eqno(78)
$$
In terms of $\chi$ we have
$$
\Pi^0\pm\Pi_3=-e^{-\frac{i\pi}4}\sqrt{2eE}\left(\frac{\partial}{\partial\chi}
\pm\frac{\chi}2\right),\quad \chi=e^{\frac{i\pi}4}T.       \eqno(79)
$$
 Taking into account also (58-59) and the relations
$$
(\Pi^0+\Pi_3)D_{\nu}(\chi)=-e^{-\frac{i\pi}4}\nu\sqrt{2eE}D_{\nu-1}(\chi),
$$
$$
(\Pi^0-\Pi_3)D_{\nu}(\chi)=e^{-\frac{i\pi}4}\sqrt{2eE}D_{\nu+1}(\chi),\eqno(80)
$$
 we find four columns of the matrix $(m-i\hat\Pi)\;{}^+Z$
 $$
\begin{bmatrix}
mD_{l-1}(\zeta)D_{-i\varkappa-1}(\chi)\\
0\\
e^{-\frac{i\pi}4}\sqrt{2eE}
D_{l-1}(\zeta)D_{-i\varkappa}(\chi)\\
-i\sqrt{2eH}D_l(\zeta)D_{-i\varkappa-1}(\chi)
  \end{bmatrix},\quad
 \begin{bmatrix}
 0\\
 mD_l(\zeta)D_{-i\varkappa}(\chi) \\
 il\sqrt{2eH}D_{l-1}(\zeta)D_{-i\varkappa}(\chi) \\
e^{\frac{i\pi}4}\varkappa\sqrt{2eE}
 D_l(\zeta)D_{-i\varkappa-1}(\chi)
  \end{bmatrix},
$$
  $$
  \begin{bmatrix}
e^{\frac{i\pi}4}\varkappa\sqrt{2eE}D_{l-1}(\zeta)D_{-i\varkappa-1}(\chi)\\
i\sqrt{2eH}D_l(\zeta)D_{-i\varkappa}(\chi)\\
mD_{l-1}(\zeta)D_{-i\varkappa}(\chi)\\
0
    \end{bmatrix},\quad
 \begin{bmatrix}
-il\sqrt{2eH}D_{l-1}(\zeta)D_{-i\varkappa-1}(\chi)\\
e^{-\frac{i\pi}4}\sqrt{2eE}D_l(\zeta)D_{-i\varkappa}(\chi)\\
0\\
mD_l(\zeta)D_{-i\varkappa-1}(\chi)
    \end{bmatrix}.                                           \eqno(81)
  $$
Here and below $e^{i(p_2x_2+p_3x_3)}$ is dropped for brevity.
We denote ${}^+\psi_1$ (${}^+\psi_2$) the fourth (first) column multiplied
by the normalization factor
$$
{}^+N_n=\exp(-\frac{\pi\varkappa}{4})
(l!2eE)^{-1/2}(eH/\pi)^{\frac14} \quad ({}^+N_n\sqrt l).  \eqno(82)
$$

Next we consider the positive-frequency solution of (73) for $t\to-\infty$
$$
{}_+Z=
{\rm diag}\{D_{l-1}(\zeta)D_{i\varkappa}(\tau),
D_l(\zeta)D_{i\varkappa-1}(\tau),D_{l-1}(\zeta)D_{i\varkappa-1}(\tau),
D_l(\zeta)D_{i\varkappa}(\tau)\}.                              \eqno(83)
$$
Here $\tau=-e^{-i\frac{\pi}4}T$. In terms of this variable we have
$$
\Pi^0\pm\Pi_3=-e^{\frac{i\pi}4}\sqrt{2eE}\left(\frac{\partial}{\partial\tau}
\mp\frac{\tau}2\right).       \eqno(84)
$$
Similarly to (80) we find
$$
(\Pi^0+\Pi_3)D_{\nu}(\tau)=e^{\frac{i\pi}4}\sqrt{2eE} D_{\nu+1}(\tau),
$$
$$
(\Pi^0-\Pi_3)D_{\nu}(\tau)=-e^{\frac{i\pi}4}\nu\sqrt{2eE}D_{\nu-1}(\tau).
                                                                   \eqno(85)
$$
With the help of these relations we get from (78) and (83) four columns
of the matrix $(m-i\hat\Pi){}_+Z$
$$
\begin{bmatrix}
mD_{l-1}(\zeta)D_{i\varkappa}(\tau)\\
0\\
e^{-\frac{i\pi}4}\varkappa\sqrt{2eE}
D_{l-1}(\zeta)D_{i\varkappa-1}(\tau)\\
-i\sqrt{2eH}D_l(\zeta)D_{i\varkappa}(\tau)
  \end{bmatrix},\quad
 \begin{bmatrix}
 0\\
 mD_l(\zeta)D_{i\varkappa-1}(\tau) \\
 il\sqrt{2eH}D_{l-1}(\zeta)D_{i\varkappa-1}(\tau) \\
e^{\frac{i\pi}4}\sqrt{2eE}
 D_l(\zeta)D_{i\varkappa}(\tau)
  \end{bmatrix},
$$
  $$
  \begin{bmatrix}
e^{\frac{i\pi}4}\sqrt{2eE}D_{l-1}(\zeta)D_{i\varkappa}(\tau)\\
i\sqrt{2eH}D_l(\zeta)D_{i\varkappa-1}(\tau)\\
mD_{l-1}(\zeta)D_{i\varkappa-1}(\tau)\\
0
    \end{bmatrix},\quad
 \begin{bmatrix}
-il\sqrt{2eH}D_{l-1}(\zeta)D_{i\varkappa}(\tau)\\
e^{-\frac{i\pi}4}\varkappa\sqrt{2eE}D_l(\zeta)D_{i\varkappa-1}(\tau)\\
0\\
mD_l(\zeta)D_{i\varkappa}(\tau)
    \end{bmatrix}.                                           \eqno(86)
$$
 Using the fourth and the first columns, we have
$$
{}_+\psi_1(x)={}_+N_n \begin{bmatrix}
-il\sqrt{2eH}D_{l-1}(\zeta)D_{i\varkappa}(\tau)\\
e^{-\frac{i\pi}4}\varkappa\sqrt{2eE}D_l(\zeta)D_{i\varkappa-1}(\tau)\\
0\\
mD_l(\zeta)D_{i\varkappa}(\tau)
    \end{bmatrix}e^{i(p_2x_2+p_3x_3)},                         \eqno(87)
$$
  $$
{}_+\psi_2(x)={}_+N_n\sqrt l\begin{bmatrix}
mD_{l-1}(\zeta)D_{i\varkappa}(\tau)\\
0\\
e^{-\frac{i\pi}4}\varkappa\sqrt{2eE}
D_{l-1}(\zeta)D_{i\varkappa-1}(\tau)\\
-i\sqrt{2eH}D_l(\zeta)D_{i\varkappa}(\tau)
  \end{bmatrix}e^{i(p_2x_2+p_3x_3)}.                         \eqno(88)
$$
Here ${}_+N_n={}^+N_n/\sqrt{\varkappa}$, see (82).

Now we note that ${}_-Z$ (${}^-Z$) can be obtained from ${}^+Z$ (${}_+Z$)
by the substitution $\chi\to-\chi$  ($\tau\to-\tau$). To obtain
 ${}_-\psi-$functions from the corresponding ${}^+\psi$-functions, we have
besides these substitution also  change the sign of $\sqrt{2eE}$ in the
columns; this is because of relations (see (79) and (80))
$$
(\Pi^0+\Pi_3)D_{\nu}(\pm\chi)=\mp e^{-\frac{i\pi}4}\nu\sqrt{2eE}
D_{\nu-1}(\pm\chi),
$$
$$
(\Pi^0-\Pi_3)D_{\nu}(\pm\chi)=\pm e^{-\frac{i\pi}4}\sqrt{2eE}
D_{\nu+1}(\pm\chi).                                         \eqno(89)
$$
Thus,
$$
{}_-\psi_1(x)={}_-N_n\begin{bmatrix}
-il\sqrt{2eH}D_{l-1}(\zeta)D_{-i\varkappa-1}(-\chi)\\
-e^{-\frac{i\pi}4}\sqrt{2eE}D_l(\zeta)D_{-i\varkappa}(-\chi)\\
0\\
mD_l(\zeta)D_{-i\varkappa-1}(-\chi)
    \end{bmatrix}e^{i(p_2x_2+p_3x_3)},                       \eqno(90)
$$
  $$
{}_-\psi_2(x)={}_-N_n\sqrt l\begin{bmatrix}
mD_{l-1}(\zeta)D_{-i\varkappa-1}(-\chi)\\
0\\
-e^{-\frac{i\pi}4}\sqrt{2eE}
D_{l-1}(\zeta)D_{-i\varkappa}(-\chi)\\
-i\sqrt{2eH}D_l(\zeta)D_{-i\varkappa-1}(-\chi)
  \end{bmatrix}e^{i(p_2x_2+p_3x_3)}, \quad {}_-N_n={}^+N_n,   \eqno(91)
$$
and similarly for ${}^-\psi_1$ and ${}^-\psi_2$.

We note by the way that the wave functions for an electron in a constant
 electric field were written down in [2] using $\gamma-$matrices in the
standard representation. Acting on these functions by an operator
$$
U=\frac1{\sqrt2}\left(\begin{array}{cc}
{\rm I}&{\rm I}\\
{\rm I}&-{\rm I}
\end{array}\right),\quad {\rm I}=\left(\begin{array}{cc}
1&0\\
0&1
\end{array}\right),
$$
we get the solutions in spinor representation. Taking into account the
magnetic field is realized by the substitutions
$$
e^{ip_2x_2}\{1, p_1-ip_2, p_1+ip_2\}\to\left(\frac{eH}{\pi}\right)^{1/4}
\frac1{\sqrt{l!}}\{D_l(\zeta), -il\sqrt{2eH}D_{l-1}(\zeta), i\sqrt{2eH}
D_{l+1}(\zeta)\}
$$
for $r=1$. For $r=2$ we have to replace in these substitutions $l$ by
$l-1$.

The electron propagator has the form
$$
G_{1/2}(x,x')=i\sum_nc_{1n}^{*-1}\left\{\begin{array}{cc}
{}^+\psi_n(x){}_+\bar\psi_n(x'),\quad t>t',\\
-{}_-\psi_n(x){}^-\bar\psi_n(x'),\quad t<t'
\end{array}\right..                                           \eqno(92)
$$
Here $\bar\psi=\psi^*\beta$ and for the constant electromagnetic field
 $n=(p_2,p_3,l,r)$, $c_{1n}^*$ is given in (27). In the spinor representation
$$
\beta=\left(\begin{array}{cc}
0&{\rm I}\\
{\rm I}&0
\end{array}\right),\quad {\rm I}=\left(\begin{array}{cc}
1&0\\
0&1
\end{array}\right),                                  \eqno(93)
$$
so that
$$
(a_1,a_2,a_3,a_4)\beta=(a_3,a_4,a_1,a_2).
$$
 Now using (81-82) and (87) we obtain
$$
{\rm tr}({}^+\psi_1(x)\;{}_+\bar\psi_1(x))=\left(\frac{eH}{\pi}\right)^{1/2}
\frac{m\exp[-\frac{\pi\varkappa}2]}{l!\sqrt{2eE\varkappa}}\times
$$
$$
D_l^2(\zeta)\{e^{-i\frac{\pi}4}
D_{-i\varkappa}(\chi)D_{-i\varkappa}(-\chi)+
e^{i\frac{\pi}4}\varkappa
D_{-i\varkappa-1}(\chi)D_{-i\varkappa-1}(-\chi)\}.       \eqno(94)
$$
Integrating over $x_1$, we get, see (46)
$$
\int\limits_{-\infty}^{\infty}dx_1
{\rm tr}({}^+\psi_1(x)\;{}_+\bar\psi_1(x))=
\frac{m\exp[-\frac{\pi\varkappa}2]}{\sqrt{2eE\varkappa}}\{e^{-i\frac{\pi}4}
D_{-i\varkappa}(\chi)D_{-i\varkappa}(-\chi)+
$$
$$
e^{i\frac{\pi}4}\varkappa
D_{-i\varkappa-1}(\chi)D_{-i\varkappa-1}(-\chi)\},
 \quad \varkappa=\frac{m^2+2eHl}{2eE},
\quad l=0,1,\cdots.                                           \eqno(95)
$$
For $r=2$ we obtain the same expression, but with $l=1,2,\cdots$

Next we multiply (95) by $i/c_{1n}^*$ according to (92) and make use of the
relation, see eq. (93) in [11] with $-i\varkappa\to-i\varkappa+1/2$
$$
\Gamma(i\varkappa)D_{-i\varkappa}(\chi)D_{-i\varkappa}(-\chi)=\sqrt2\int
\limits_0^{\infty}\frac{d\theta}{\sqrt{\sinh2\theta}}e^{-i2\varkappa\theta+
\theta -\frac i2T^2\tanh\theta},\quad \theta=eEs,     \eqno(96)
$$
and relation obtainable from this by the substitution
 $i\varkappa\to i\varkappa+1$.

 Now we get from (95-96)
$$
\int\limits_{-\infty}^{\infty}dx_1\frac{i}{c_{1n}^*}
{\rm tr}({}^+\psi_1(x)\;{}_+\bar\psi_1(x))=-\frac{m\exp(-i\frac{\pi}4)}
{\sqrt{2\pi eE}}\times
$$
$$
\int
\limits_0^{\infty}\frac{d\theta}{\sqrt{\sinh2\theta}}2\cosh\theta
e^{-i2\varkappa\theta-
\frac i2T^2\tanh\theta}, \quad T=\sqrt{2eE}\left(t-\frac{p_3}{eE}\right).
                                                                    \eqno(97)
$$
Integrating this expression over $t$, we obtain
$$
\int\limits_{-\infty}^{\infty}dt\int\limits_{-\infty}^{\infty}dx_1
\frac{i}{c_{1n}^*}
{\rm tr}({}^+\psi_1(x)\;{}_+\bar\psi_1(x))=
$$
$$
im\int\limits_0^{\infty}ds\coth (eEs)
e^{-is(m^2+2eHl)},\quad l=0,1,2,\cdots
                                                                    \eqno(98)
$$
 For $r=2$ we have the same expression, but with $l=1,2,\cdots$.

Taking into account the remarks after eq. (70a) and inserting the
regularizer $R(\theta,\tau)$ in the integrand, we obtain the contribution
to ${\cal L}_{1/2}$ from the state $n=(p_2,p_3,l,r)$.
Summing up over $l$ and $r$ (see (40)) and integrating over
 $\frac{dp_2}{2\pi}$ and
$\frac{dp_3}{2\pi}$ (see (39) and (12)), we obtain in agreement with (41)
$$
{\cal L}_{1/2}=\frac{e^2HE}{8\pi^2}\int\limits_0^{\infty}\frac{ds}se^{-ism^2}
\coth\theta\cot\tau R(\theta,\tau).           \eqno(99)
$$

Finally we note that for $H=0$ we have instead of (98)
$$
\int\limits_{-\infty}^{\infty}dt
\frac{i}{c_{1n}^*}
{\rm tr}({}_+\psi_1(x)\;{}^+\bar\psi_1(x))=
$$
$$
im\int\limits_0^{\infty}ds\coth eEs
e^{-is(m^2+p_1^2+p_2^2)},\quad l=0,1,2,\cdots                  \eqno(100)
$$
Inserting $1/(2ms)$ and $R(\theta,\tau)$, we get the agreement with (35).

\section{Vector boson}
The propagator and the effective Lagrange function for the vector boson
with gyromagnetic ratio $g=2$ in a constant electromagnetic field was
obtained by Vanyashin and Terentyev [17]. In another form the
propagator was found by the author [16]. In this latter paper there is
a misprint in eq. (73), where the argument of $\sin$ and $\cos$ should be
$2\tau$ not $\tau$. Besides, the statement, that the divergence term in the
expression for the current in eq. (38) does not contribute, is not true
when the magnetic field is present; this, however, is of no consequence
since the expression was used only for the normalization of wave functions.

From the results of Vanyashin and Terentyev it follows that the relation (48)
holds also for the vector boson, if we mean that $G_b=G^{\mu}{}_{\mu}$.
 Using (48) we can reproduce the expression for
${\cal L}_1$ in [17] starting from our propagator. Indeed, our result for
$$
G^{\mu}{}_{\mu}=\frac{e^2EH}{16\pi^2}\int_C\frac{ds}{\sinh\theta
\sin\tau }e^{-ism^2}\times
$$
$$
\{2\cos2\tau+2\cosh2\theta-\frac{i}{m^2}[eH\cot\tau
+eE\coth\theta]\}                                       \eqno(101)
$$
can be written in a more simple form, if we note that
$$
\frac{d}{ds}\frac1{\sinh\theta \sin\tau}=-\frac1{\sinh\theta \sin\tau}
[eH\cot\tau +eE\coth\theta]\},\quad \tau=eHs,\quad
\theta=eEs.     \eqno(102)
$$
Then we can integrate by parts the term in the square brackets in (101)
$$
-\frac{ie^2EH}{16\pi^2m^2}\int_C\frac{ds}{\sinh\theta
\sin\tau }e^{-ism^2}[eH\cot\tau+eE\coth\theta]\} \Longrightarrow
$$
$$
-\frac{e^2EH}{16\pi^2}\int_C\frac{ds}{\sinh\theta
\sin\tau }e^{-ism^2}                          \eqno(103)
$$
 Here we discarded divergent term independent of $E$ and $H$. So (101) is
equivalent to
$$
\frac{e^2EH}{16\pi^2}\int_C\frac{ds}{\sinh\theta\sin\tau }e^{-ism^2}
\{2\cos2\tau+2\cosh2\theta-1\}.                                 \eqno(104)
$$
Inserting $(-1/s)$ in the integrand, we get eq. (21) in [17] and we agree
with the subsequent formulas in that paper.

Returning to our present problem we note that for the constant electric
field $c_{1n}^*$ is independent of the polarization state of vector boson
and is the same as in the scalar case [11]. Nevertheless ${\rm Im}{\cal L}_1$
is not simply $3{\rm Im}{\cal L}_0$ [17]. Thus, the knowledge of $c_{1n}^*$
is of no use in obtaining $\ln C_{1n}^{*ren}$. Resorting to the proper-time
method, we find that the  problem is more difficult then in the previous
cases. As seen already from (101), the dependence on $m^2$ is more
complicated here and the contributions from the electric and magnetic fields
are not factorized in the proper-time integrand. For these reasons we consider
here only the constant magnetic field.

From [11] we have for the spin states $r=1, 2, 3$
$$
{}_+\psi_1^{\mu}(x)\;{}_+\psi_{1\mu}^*(x)=
$$
$$
\left(\frac{eH}{\pi}\right)^{1/2}
\frac1{2|p^0|l!}\frac1{(l+1)(m^2+eHl)}\{-(l+1)^2eHD_l^2(\zeta)
+[m^2+eH(2l+1)]D_{l+1}^2(\zeta)\};                    \eqno(105)
$$
  $$
{}_+\psi_2^{\mu}(x)\;{}_+\psi_{2\mu}^*(x)=\left(\frac{eH}{\pi}\right)^{1/2}
\frac1{2|p^0|l!}D_l^2(\zeta);                                \eqno(106)
  $$
$$
{}_+\psi_3^{\mu}(x)\;{}_+\psi_{3\mu}^*(x)=
$$
$$
\left(\frac{eH}{\pi}\right)^{1/2}
\frac1{2|p^0|l!}\frac l{2m^2(m^2+eHl)}
\{-2eH[m^2+eH(2l+1)]D_l^2(\zeta)+
+[eHD_{l+1}(\zeta)
$$
$$
-(m^2+eHl)D_{l-1}(\zeta)]^2+[eHD_{l+1}(\zeta)+
(m^2+eHl)D_{l-1}(\zeta)]^2]\}.                    \eqno(107)
$$
Integrating with the help of (46) the expressions in (105-107) over $x_1$,
we get in all three cases $1/(2|p^0|)$, but
$$
l=l_{min}, l_{min}+1, \cdots,\quad l_{min}=\left\{\begin{array}{ccc}
-1,\quad r=1\\
0,\quad r=2\\
1,\quad r=3.
\end{array}\right.                \eqno(108)
$$
The vector boson propagator has the form [11]
$$
G_1^{\mu\nu}(x,x')=i\int\limits_{-\infty}^{\infty}\frac{dp_2}{2\pi}
 \int\limits_{-\infty}^{\infty}\frac{dp_3}{2\pi}
\sum_{r=1}^3\sum_{l_{min}}^{\infty}
\left\{\begin{array}{cc}
{}_+\psi_n(x)^{\mu}{}_+\psi_n^{*\nu}(x'),\quad t>t',\\
{}_-\psi_n^(x)^{\mu}{}_-\psi_n^{*\nu}(x'),\quad t<t'
\end{array}\right.                               \eqno(109)
$$
We see from (68), (48) and the above results that the contribution to
${\cal L}_1$
from the state with quantum numbers $n=(p_2,p_3,l,r)$ is
$$
i\ln c_{1n}^*=-i\frac{e^{\frac{i\pi}4}}{2\sqrt{\pi}}
\int\limits_0^{\infty}
\frac{ds}{(s)^{3/2}}e^{-is[m^2+eH(2l+1)+p_3^2]}.                 \eqno(110)
$$
The sum over $r$ and $l$ is performed with the help of formula obtainable
from (25)
$$
\sum_{r=1}^3\sum_{l_{min}}^{\infty}e^{-iseH(2l+1)}=\frac1{2i\sin eHs}[1+
2\cos 2eHs].                                                  \eqno(111)
$$
  To integrate over $\frac{dp_2}{2\pi}$  and $\frac{dp_3}{2\pi}$, we use
(39) and (12). Then, inserting $R(\tau)$, we obtain
$$
\sum_nW_{spin1,n}=-\frac{eH}{16\pi^2}
\int\limits_0^{\infty}
\frac{ds}{s^2\sin{\tau}}e^{-ism^2}[3-4\sin^2\tau]R(\tau).      \eqno(112)
$$
 $R(\tau)$ is defined in accordance with paper [17]
$$
\frac{3-4\sin^2\tau}{\sin\tau}\to3\left(\frac1{\sin\tau}-
\frac1{\tau}-\frac{\tau}6\right)-4(\sin\tau-\tau)=
\frac{3-4\sin^2\tau}{\sin\tau}R(\tau).         \eqno(113)
$$
From here
$$
R(\tau)=1-\frac{\sin\tau}{3-\sin^2\tau}\left(\frac3{\tau}-\frac72\tau\right),
\quad R(\tau)|_{\tau\ll1}=\frac{29}{120}\tau^4.             \eqno(114)
$$
Hence from (110) we have
$$
i\ln C_{1n}^{*ren}=-i\frac{e^{\frac{i\pi}4}}{2\sqrt{\pi}}
\int\limits_0^{\infty}
\frac{ds}{(s)^{3/2}}e^{-is[m^2+eH(2l+1)+p_3^2]}R(\tau).           \eqno(115)
$$
$l$ are given in (108). Substituting $\tau\to-it$ and rotating the
integration contour, we see that $\ln C_{1n}^{*ren}$ is real as it should be
for the magnetic field.

\section{Conclusion}
We have shown how the renormalized phase of vacuum-vacuum amplitude in
Quantum Electrodynamics can be expressed through the properly fixed phases of 
Bogoliubov coefficients; the nonzero phase of the former indicates the 
nonzero phases of the latter.  In general the  knowledge of Bogoliubov 
coefficients is insufficient for obtaining the phase of vacuum-vacuum 
amplitude. Additional information is needed. Thus, in the case of the
constant magnetic and constant electromagnetic  fields we have used 
the symmetry between the electric and magnetic fields in the Lagrange
function. In the case of vector boson the knowledge of Bogoliubov
coefficients is of no use for fixing its phases. The resort to the
 proper-time method shows that the expressions for phases are in general
 more complicated then for lower spin particles. For this reason we have
 presented the results only for the constant magnetic field, where they
 turned out as simple as expected.

\section{Acknowledgments}
I am grateful to V.I.Ritus for discussions which led to the appearance
of this paper. This work was supported in part by the Russian Foundation for
Basic Research (projects no. 00-15-96566 and 02-02-16944).
\section*{References}
\begin{enumerate}
\item  N.B.Narozhny and A.I.Nikishov, Yad. Fiz. {\bf 11}, 1072 (1970). \\
\item A.I.Nikishov, Tr. Fiz. Inst. Akad. Nauk SSSR {\bf 111}, 152 (1979);
 J. Sov.
 Laser Res. {\bf 6},
 619 (1985). \\
 \item A.A.Grib, S.G.Mamaev, and V.M.Mostepanenko,{\sl Vacuum Quantum Effects
  in Strong Fields} (Energoatomizdat, Moscow, 1988).\\
\item A.I.Achiezer, V.B.Berestetskii, {\sl Quantum Electrodynamics}
(Moscow,1969)\\
\item V.B.Berestetskii, E.M.Lifshits and L.P.Pitaevskii
{\sl Quantum Electrodynamics}(Nauka,Moscow,1989; Pergamon, Oxford, 1982)\\
\item W.Heisenberg, H. Euler, Zs.f.Phys. {\bf 98}, 714 (1936).\\
\item V.Weisskopf, Kgl, Dan. Vidensk. Selsk. Mat-Fys. Medd., {\bf 14},
 No.6 (1936).\\
\item J.Schwinger, Phys. Rev. {\bf 82}, 664 (1951).   \\
\item V.I.Ritus, ZhETF {\bf69}, 1517 (1975); ZhETF {\bf73}, 807 (1977);
Tr. Fiz. Inst. Akad. Nauk SSSR {\bf 168}, 52 (1986); in
{\sl Issues in Intense-Field Quantum Electrodynamics}, Ed. by V.L.Ginzburg
(Nova Science, Commack, 1987).\\
\item B.S.De Witt, Physics Reports, {\bf 19c}, 227 (1975).\\
 \item A.I.Nikishov, ZhETF, {\bf 120}, 227 (2001).\\
\item Y.L.Luke, {\sl Mathematical functions and their approximations},
Academic Press Inc. 1975.\\
\item R.B.Dingle, {\sl Asymptotic Expansions: Their Derivation and
 Interpretation},
Academic Press, London and New York, 1973.\\
\item I.S.Gradstein, I.M.Ryzhik, {\sl Tables of Integrals, Sums, Series,
    and Products }, Moscow, 1962.                    \\
\item M.Abramowits and I.Stegun, {\sl Handbook of Mathematical Functions},
National Bureau of Standards, Applied Mathematical Series, 55, Washington, 1964.                       \\
\item A.P.Prudnikov, Yu. A.Brychkov, and O.I.Marichev, {\sl Integrals and
    Series. Special Functions}, Moscow, Nauka, 1983.
 \item V.S.Vanyashin and M.V.Terentyev, Zh. Eksp. Teor. Fiz. {\bf 48}, 565
(1965) [Sov.Phys. JETP {\bf 21},375 (1965)].\\
 \end{enumerate}
\end{document}